\documentstyle[12pt]{article}
\begin{document}
%
%
\def\R{\rm l\!R\,}
\def\Compton{\overline{\phantom{\overline{.\!.}}}\!\!\!\lambda}

\title{\bf Magnetic permeability of constrained scalar QED vacuum}

\author{M. V. Cougo-Pinto\thanks{e-mail:{\it marcus@if.ufrj.br}}, 
C. Farina\thanks{e-mail:{\it farina@if.ufrj.br}}, 
M. R. Negr\~ao \thanks{e-mail:{\it guida@if.ufrj.br}} \\ 
and \\ A. C. Tort\thanks{e-mail:{\it tort@if.ufrj.br}}\\
\\{\small\it Instituto de F\'{\i}sica, 
Universidade Federal do Rio de Janeiro}\\
{\small\it Caixa Postal 68528, Rio de Janeiro, RJ 21945-970, Brazil}}
\date{\today}
\maketitle
\begin{abstract}
{\scriptsize We compute the influence of boundary conditions on the 
Euler-Heisenberg effective 
Lagrangian scalar QED scalar for the case of a pure  
magnetic field. The boundary conditions constrain the quantum scalar field 
to vanish on two parallel planes separeted by a distance $a$ 
and the magnetic field is assumed
to be constant, uniform and perpendicular to the planes. The effective Lagrangian
is obtained using Schwinger's  proper-time representation and exhibits 
new contributions generated by the boundary condition much in the same way as
a material pressed between two plates exhibits new magnetic properties. The 
confined bosonic vacuum presents the expected diamagnetic properties and besides
the new non-linear $a$-dependent contributions to the 
susceptibility we show that there exists also a new $a$-dependent 
contribution for the vacuum  permeability in the linear approximation.}
\end{abstract}
\vskip 1 true cm

The classical electromagnetic field $F^{\mu\nu}$ in classical vacuum is 
described by the Maxwell Lagrangian $-F^{\mu\nu}F_{\mu\nu}/4$. We consider
here the case in which there is only a magnetic field ${\bf B}$ and so the 
Lagrangian is given by:
\begin{equation}\label{Maxwell}
{\cal L}^{(0)}({\bf B})=-\frac{1}{2}{\bf B}^2 \; .
\end{equation}
If the field is in a medium it can be described by an effective Lagrangian 
${\cal L}^{(0)}({\bf B})+{\cal L}^{(1)}({\bf B})$, where 
${\cal L}^{(1)}({\bf B})$ takes in account the net influence of the 
medium on the field
by means of permeability constants and functions. The permeability constant
appears in the quadratic term of ${\cal L}^{(1)}({\bf B})$ while the higher 
order terms of ${\cal L}^{(1)}({\bf B})$ describe non-linear 
susceptibility effects. In their seminal work Euler and
Heisenberg noticed that usual QED vacuum behaves as a medium under an applied
electromagnetic field and obtained for this vacuum the one-loop effective 
Lagrangian \cite{H-E,Schwinger51}. An effective Lagrangian can  
be obtained also for the bosonic charged vacuum of scalar QED \cite{Schwinger51}
which is the case that interests us here.
The idea of effective Lagrangians is now an important tool in quantum field 
theory and in the original case of QED, which provides an interesting and not 
too complicated formalism, it has a privileged setting to test new methods and 
ideas \cite{DittrichReutter85}. There is also an ongoing experimental effort
to measure non-linear effects under strong magnetic fields predicted by 
the Euler-Heisenberg Lagrangian
\cite{IacopiniZavattini79,Cantatore-etal91,PVLAScolaboration97} and 
recent experiments involving intense electron and laser beams have shown
that the critical electrical field strength of the Euler-Heisenberg Lagrangian 
may be within reach \cite{Burke97,Melissinos98}.

Vacuum fluctuations of any field are affected by 
the imposition of boundary conditions, a phenomenum generally known as Casimir 
effect \cite{MostepanenkoTrunovB,PlunienMullerGreinerR}. 
In the original Casimir
effect \cite{Casimir48} two large metallic parallel plates implement
boundary conditions on the  vacuum fluctuations of the quantum electromagnetic 
field. The influence of the boundary conditions appears as a force of 
attraction between the plates. This force 
was measured by Sparnaay \cite{Sparnaay58} and more recently with high 
precision in experiments
conducted by Lamoreaux \cite{Lamoreaux97} and by Mohideen and Roy 
\cite{MohideenRoy98}. In the case of a charged quantum field the vacuum 
fluctuations can be influenced by an applied external field and by the 
imposition of boundary conditions and it poses by itself the question of 
how those two agents together affect the charged vacuum fluctuations. 
There are two quite distinct points of view in which to pose the question
and they 
lead to different physical phenomena to be investigated. From one
point of view
we ask how the boundary conditions affects the Euler-Heisenberg effective 
Lagrangian and from the other how the external applied field affects the 
Casimir energy. In a previous work we have considered the influence of 
boundary conditions on the effective Lagrangian of a charged fermionic vacuum 
\cite{C-PFRT98}. Here we wish to consider the influence of the 
boundary conditions on a charged bosonic vacuum. 
In both situations the obtained results are in 
accord with the general view that fermionic and bosonic vacua should exhibit
paramagnetic and diamagnetic properties, respectively. 
Let us notice that we can think of 
the boundary conditions as providing an applied stress on the slab of 
vacuum between the plates and from this point of view it is quite natural
to have a change in the vacuum constitutive relations when the separation
of the plates $a$ is changed. 
We should mention that the refractive index of a non-trivial
vacuum can be related with the vacuum expectation value of the stress tensor
in a general formalism in the framework of QED effective actions 
\cite{DittrichGies98}. We have also considered
the other point of view and studied 
the influence of an external magnetic field on the Casimir energy of bosonic
\cite{C-PFNT98} and fermionic \cite{C-PFT98} fields. Our results show that 
the external field enhances the fermionic Casimir effect and inhibits the
bosonic Casimir effect. 

The question of how boundary conditions affect the 
effective Lagrangian is an important one because vacuum fluctuations can
be constrained by boundary conditions implemented by several types of cavities
and in the case of QCD the confining boundary conditions on vacuum fluctuations
cannot be avoided.
Certainly the QCD problem is immensely complicated and by this very reason
it renders useful a previous investigation in 
a simpler context, one in which we can fix our
attention on the main features of the problem before facing more 
complicated gauge groups and boundary conditions. Therefore, let us consider
here the simple situation of a massive charged scalar field 
between two parallel planes on which we assume Dirichlet boundary conditions
for the field. We make our calculation
considering the planes as large square plates with 
side $\ell$ and separated by a distance $a$ ($\ell\gg a$); at the 
appropriated moment the limit $\ell\rightarrow\infty$ may be taken. We 
consider the applied external magnetic field ${\bf B}$ as constant and 
uniform which is perpendicular to the plates and points in a direction
that makes $eB$ positive. With the 
charged scalar field we are free of kinematical complexities, the 
choice of a pure magnetic external field excludes the possibility 
of pair creation for any field strength, and the Dirichlet boundary conditions
on the plates is a rather simple constraint on the charged vacuum which changes
its magnetic permeability properties. Let us observe that the permeability 
and permittivity properties of QED vacuum may also be changed by considering
boundary conditions imposed on the quantum electromagnetic field while the
quantum charged field is free of boundary conditions. In this case the change in
the vacuum properties requires a two-loop process and it is known as Scharnhorst
effect \cite{Scharnhorst90,Barton90,BartonScharnhorst93}
(see also \cite{C-PFST98a,C-PFST98b}). On the other hand we study here 
boundary conditions on the charged quantum
field and the resulting changes in the vacuum properties appear
already at the one-loop level.

To obtain the effective Lagrangian of the charged scalar fluctuations in 
the presence of the magnetic field we will employ Schwinger's 
proper time representation for 
the effective action \cite{Schwinger51} and a straightforward method of 
calculation also used by Schwinger to calculate the 
original Casimir energy \cite{Schwinger92} and 
later applied to several other problems 
\cite{C-PFRT98,C-PFNT98,C-PFT98,C-PFSST}. The
proper time representation for the effective action ${\cal W}^{(1)}$
is given by \cite{Schwinger51}:
\begin{eqnarray}
{\cal W}^{(1)} = - \frac{i}{2} \int_{s_0}^{\infty}
\frac{ds}{s} Tr e^{-isH},
\label{W}
\end{eqnarray}
where $s_{0}$ is a cutoff in the proper-time $s$, $Tr$ stands for the total
trace and $H$ is the proper-time Hamiltonian, which in the present case of 
a scalar field is given by $(p -
eA)^{2} + m^{2}$, where $p_{\mu} = - i {\partial}_{\mu}$, $m$ is the mass 
of the scalar field, $e$ is its charge and $A$ is the potential of the 
external magnetic field ${\bf B}$, whose direction can be arbitrarily chosen to
make $eB$ positive. The spatial components of
$p$ parallel to the plates are constrained by the Landau levels created
by ${\bf B}$ with multiplicity $eB\;\ell^2/2\pi$ and the component of $p$ 
perpendicular to the plates is discretized by Dirichlet boundary
conditions into the values $\pi n/a$, where $n$ is a positive integer. 
The time component of $p$ has as eigenvalue any real number $\omega$ and
the charge degrees of freedom of the complex scalar field contributes
with a factor of 2 to the total trace. Therefore, we obtain for the trace in
(\ref{W}):
\begin{eqnarray}
Tr e^{-isH} = 2 e^{-ism^{2}}\sum_{n'=0}^\infty
\frac{eB{\ell}^{2}}{2\pi}
e^{-iseB(2n'+1)} \sum_{n=1}^\infty e^{-is(n\pi
/a)^{2}} \int \frac{dtd\omega}{2\pi}
e^{is\omega^{2}} \; ,
\label{traco1}
\end{eqnarray}
where the range of the time integral is taken to be a large observation
time $T$. For the first sum in (\ref{traco1}) we have
\begin{eqnarray}
\sum_{n'=0}^\infty \frac{eB{\ell}^{2}}{2\pi}
e^{-iseB(2n'+1)} = 
\frac{{\ell}^2}{4\pi is}
[1+iseB\,{\cal M}(iseB)] \; ,
\label{Landau}
\end{eqnarray}
where ${\cal M}$ is the function defined by:
\begin{eqnarray}
{\cal M}(\xi) = cosech{\xi} - {\xi}^{-1}.
\label{myfunc}
\end{eqnarray}
The function ${\cal M}$ plays in the present bosonic formalism the same role 
as the Langevin function in the fermionic formalism \cite{C-PFRT98}.
For the second sum we make use of Poisson sum formula 
\cite{Poisson1823} in order to obtain:
\begin{eqnarray}
\sum_{n=1}^\infty e^{-is(n\pi/a)^{2}} =
\frac{a}{\sqrt{i\pi s}}
\sum_{n=1}^\infty e^{i(an)^{2}/s} +
\frac{a}{2\sqrt{i\pi s}} - \frac{1}{2}.
\label{Poisson}
\end{eqnarray}
Using (\ref{Poisson}) and (\ref{Landau}) into (\ref{traco1}), we obtain for 
the trace:
\begin{eqnarray}\nonumber
Tr e^{-isH} = \frac{a{\ell}^{2}T}{4{\pi}^{2}}&&\frac{e^{-ism^{2}}}{is}
[1+iseB{\cal M}(iseB)] \times\\
&&\times\biggl[ \frac{1}{2\sqrt{i\pi s}}
+ \frac{1}{\sqrt{i\pi s}}\sum_{n=1}^\infty e^{i(an)^{2}/s} 
-\frac{1}{2a}
\biggl]\sqrt{\frac{\pi}{-is}}\; .
\label{traco2}
\end{eqnarray}
Now we substitute equation (\ref{traco2}) into the effective action 
(\ref{W}) and following Schwinger \cite{Schwinger51,Schwinger92} we 
use Cauchy contour theorem to make a $\pi/2$ clockwise rotation in 
the integration $s$-axes in order to obtain the effective action 
as the following manifestly real quantity:
\begin{eqnarray}\nonumber
&&{\cal W}^{(1)}=
\frac{a\ell^2T}{16\pi^2}
\int_{s_0}^{\infty}
\frac{ds}{s^3} e^{-sm^2} [1+seB\,{\cal M}(seB)]+\\
&+&\frac{a\ell^2T}{8\pi^2}
\int_{s_0}^\infty\frac{ds}{s^{3}} e^{-sm^2} 
\biggl [\sum_{n=1}^\infty e^{-(an)^{2}/s}+\frac{\sqrt{\pi s}}{2a} \biggl ]\, 
[ 1+ seB{\cal M}(seB)] \; .
\label{L^(1)}
\end{eqnarray}
From this action we obtain the corresponding effective Lagrangian
which we add to the Maxwell Lagrangian (\ref{Maxwell}) to obtain the 
still unrenormalized cutoff dependent complete Lagrangian:
\begin{eqnarray}\nonumber
&&{\cal L}(a,B) =-\frac{1}{2}B^2 
+\frac{1}{16\pi^2}
\int_{s_0}^{\infty}
\frac{ds}{s^3} e^{-sm^2} [1+seB\,{\cal M}(seB)]+\\
&+&\frac{1}{8\pi^2}
\int_{s_0}^\infty\frac{ds}{s^{3}} e^{-sm^2} 
\biggl [\frac{\sqrt{\pi s}}{2a} +
\sum_{n=1}^\infty e^{-(an)^{2}/s} \biggl ]\, [ 1+ seB\,{\cal M}(seB)] \; .
\label{L}
\end{eqnarray}
Now we follow the usual procedure for renormalizing this Lagrangian
\cite{H-E,Schwinger51,DittrichReutter85}. 
Consider in the first integral in equation (\ref{L}) the first two 
terms in the expansion of the function $1+$$seB{\cal M}(seB)$ 
in powers of $seB$, to wit: $1$ and $(seB)^3/6$.
The first term gives for the first integral in (\ref{L}) a contribution
which in the limit $s_0\rightarrow 0$ tends to the infinite constant 
$m^4\Gamma(-2)/16 \pi^2$, where $\Gamma$ is the Euler 
gamma function. This infinity can be eliminated by simply 
subtracting it from the Lagrangian ${\cal L}(a,B)$. The second 
term gives for the first integral in 
(\ref{L^(1)}) a contribution which is proportional to $B^2/2$,
with a constant of proportionality  which tends to the infinite constant 
$-e^2\Gamma(0)/48 \pi^2$ in the limit $s_0\rightarrow 0$. We write this
constant as $Z_3^{-1}-1$ and absorb it in the definitions of renormalized
charge, $e_R=eZ_3^{1/2}$, and renormalized field, $B_R=BZ_3^{-1/2}$. With this 
procedure the second infinity is also disposed off, in the first term 
in (\ref{L}) the bare field $B$ is replaced by $B_R$ and in the rest of 
${\cal L}(a,B)$
we have $eB$ replaced by $e_RB_R$. Notice that the rest of ${\cal L}(a,B)$
does not change its form under these replacements because $e_RB_R=eB$.
After the introduction of these renormalization procedures 
all the infinities are removed from the Lagrangian ${\cal L}(a,B)$ 
and we can take the limit $s_0\rightarrow 0$
in order to eliminate the cutoff. The first term in the expansion
of $1+$$seB{\cal M}(seB)$ gives for the second integral in (\ref{L}) 
a finite term which is actually the Casimir energy of the charged
scalar field \cite{PlunienMullerGreinerR}. However, this energy is totally 
irrelevant for the 
effective Lagrangian because is independent of the field $B$.
Therefore we can for convenience also subtract this term from 
${\cal L}(a,B)$ to 
obtain a final well-defined renormalized Lagrangian ${\cal L}_R$ written
in terms of the renormalized quantities $e_R$ and $B_R$. With the proviso
that from now on we will deal only with renormalized quantities we
write the final renormalized form of the effective Lagrangian 
suppressing all subindexes $R$, thereby obtaining:
\begin{equation}\label{L_R}
{\cal L}(a,B) = -\frac{1}{2} B^2+{\cal L}_{HE}^{(1)}(B)+
{\cal L}_{HEC}^{(1)}(a,B) \; ,
\end{equation}
where
\begin{eqnarray}
{\cal L}_{HE}^{(1)}(B) = \frac{1}{16\pi^2}
\int_0^{\infty}
\frac{ds}{s^3} e^{-sm^2} [seB\,{\cal M}(seB)+\frac{1}{6}(seB)^2]
\label{HE}
\end{eqnarray}
and
\begin{equation}
{\cal L}_{HEC}^{(1)}(a,B) =
\frac{1}{8\pi^2}
\int_0^\infty\frac{ds}{s^{3}} e^{-sm^2} 
\biggl [\frac{\sqrt{\pi s}}{2a} +
\sum_{n=1}^\infty e^{-(an)^{2}/s} \biggr ]\,seB\,{\cal M}(seB) \; .
\label{HEC}
\end{equation}
The result in (\ref{HE}) agrees with the effective Lagrangian
first obtained by Schwinger \cite{Schwinger51} which is 
the version for the charged scalar field of the
effective Lagrangian obtained by Euler and Heisenberg for the 
charged Dirac field\cite{H-E}.
The expression for ${\cal L}_{HE}^{(1)}(B)$ in (\ref{HE}) can be 
further worked out
to express the integral in it in terms of a Riemann zeta function 
\cite{Dittrich76}. Let us consider the regime of $B$ small
compared to the critical field $B_{cr}=m^2/e$. In this regime
${\cal L}_{HE}^{(1)}(B)$ can be expanded in powers of $B^2$ and we obtain
(cf. formula {\bf 1411},12 in \cite{GradshteijnRyzhik65}):
\begin{equation}\label{HEexpansion}
{\cal L}_{HE}^{(1)}(B)=-\frac{m^4}{16\pi^2}\sum_{k=2}^\infty 
\frac{(2^{2k-1}-1)B_{2k}}{k(2k-1)(2k-2)} 
\left(\frac{B}{B_{cr}}\right)^{2k} \; ,
\end{equation}
where  $B_{2k}$  is the $2k$-th Bernoulli number. This expansion shows
that the lowest order contribution from the Euler-Heisenberg effective 
Lagrangian is a term in $B^4$, which means a non-linear contribution
to the magnetic susceptibility. If we want this contribution 
to be relevant
the external field $B$ should not be much smaller than the critical field
$B_{cr}$. 
The effective Lagrangian ${\cal L}_{HEC}^{(1)}(a,B)$ 
in (\ref{HEC}) is the result that we were looking for. It takes into 
account the effect of the boundary conditions on the charged vacuum 
fluctuations and can be called the Casimir-Euler-Heisenberg 
effective Lagrangian for the charged scalar field. 

The result that we have obtained 
in (\ref{HEC}) for the bosonic vacuum and the previous 
result for the fermionic vacuum \cite{C-PFRT98} shows 
quantitatively how the magnetic permeability of both vacua changes with the
parameter $a$ of the boundary conditions. In the limit
$a\rightarrow\infty$ we have that $ {\cal L}_{HEC}^{(1)}(a,B)\rightarrow 0$
and (\ref{L_R}) reduces to the usual effective Lagrangian (\ref{HE}) 
of the unconstrained charged scalar vacuum. For finite $a$ we expect from
(\ref{HEC}) significant contributions to the complete effective Lagrangian if
the dimensionless parameter $am$ is not large. To see more clearly the 
new contributions yielded by the Casimir-Euler-Heisenberg effective Lagrangian
(\ref{HEC}) to the permeability properties of vacuum let us rewrite it 
in the following form:
\begin{equation}
{\cal L}_{HEC}^{(1)}(a,B) =-\frac{1}{2}\biggl[ \frac{1}{\mu(am)}-1\biggr]
+{\cal L}_{HE}^{(1)\;\prime}(a,B) \; ,
\label{mu-term+HECprime}
\end{equation}
where
\begin{equation}\label{mu}
\frac{1}{\mu(am)}=1+\frac{e^2}{48\pi\,am}+\frac{e^2}{12\pi^2}
\sum_{n=1}^\infty K_0(2amn) \; ,
\end{equation}
and
\begin{equation}
{\cal L}_{HE}^{(1)\;\prime}(a,B) =
\frac{1}{8\pi^2}
\int_0^\infty\frac{ds}{s^{3}} e^{-sm^2} 
\biggl [\frac{\sqrt{\pi s}}{2a} -
\sum_{n=1}^\infty e^{-(an)^{2}/s} \biggr ]\,\biggl[seB\,{\cal M}(seB)
+\frac{1}{6}(seB)^2\biggr] \; .
\label{HEprime}
\end{equation}
The expression in (\ref{HEprime}) provides the new non-linear 
$a$-dependent contributions
to the magnetic susceptibility stemming jointly from the external field $B$
and the boundary conditions. It provides at each order in the expansion
(\ref{HEexpansion}) a contribution due to the influence of the boundary 
conditions. Those contributions are given explicitly by the following
expansion of ${\cal L}_{HE}^{(1)\;\prime}(a,B)$  in the regime of small
$B$ (cf. formula {\bf 1411},12 in \cite{GradshteijnRyzhik65}):
\begin{eqnarray}\label{HEprimeexpansion}
\nonumber
{\cal L}_{HE}^{(1)\,\prime}(B)=&-&\frac{m^4}{16\pi^2}\sum_{k=2}^\infty 
\frac{(2^{2k-1}-1)B_{2k}}{k(2k-1)(2k-2)}\times\\
&\times&\biggr[ \frac{4\pi}{3\,am}+2\sum_{n=1}^\infty \frac{(2amn)^{2k-2}}{(4k-5)!!}
K_{2k-2}(2amn)\biggr]
\left(\frac{B}{B_{cr}}\right)^{2k} \; .
\end{eqnarray}
In this expression we can read at each order of $B/B_{cr}$ the corrections
to be added to (\ref{HEexpansion}) if the boundary conditions are enforced.

In (\ref{mu}) we have from ${\cal L}_{HEC}^{(1)}(a,B)$ a type of 
contribution to the permeability constant which is not present in the
Euler-Heisenberg Lagrangian (\ref{HE}). This novel contribution provides
the charged scalar vacuum with an $a$-dependent permeability constant $\mu(am)$. From the
properties of the $K_0$ Bessel function \cite{GradshteijnRyzhik65}
we see in (\ref{mu}) that
$\mu(am)$ is striclty less than 1, {\it i.e.}, the charged scalar vacuum
magnetic permeability is definitely diamagnetic. We also see that the 
expected limit of $\mu(am)\rightarrow 1$ is obtained when 
$a\rightarrow\infty$ and that $\mu(am)\rightarrow 0$ when 
$a\rightarrow 0$. From those two limits the general behaviour
of the permeability constant is quite predictable on physical grounds 
and is depicted in Figure 1. To highlight the importance of the permeability
constant $\mu(am)$ generated by the boundary conditions let us consider
the week field regime in which only quadratic terms in the Lagrangian are
not negligible. In this situation the whole Euler-Heisenberg Lagrangian
${\cal L}_{HE}^{(1)}(B)$ (\ref{HE}) as well as the corrections 
${\cal L}_{HE}^{(1)\;\prime}(a,B)$ in (\ref{HEprime}) make no contribution 
to the effective Lagrangian  and (\ref{L_R}) reduces to the expression:
\begin{equation}\label{-B^2/2mu}
{\cal L}(a,B) = -\frac{1}{2}\frac{B^2}{\mu(am)} \; ,
\end{equation}
in which the charged vacuum manifests itself only through the permeability
constant $\mu(am)$.  
 
In this paper we have obtained the effect on the Euler-Heisenberg 
effective Lagrangian
of the charged scalar field due to Dirichlet boundary conditions on two 
parallel plates. We have obtained corrections to the non-linear 
susceptibility effects of the Euler-Heisenberg Lagrangian and also
a novel contribution providing the charged scalar vacuum with a 
permeability constant which depends on the boundary conditions. 
The permeability constant shows that the bosonic vacuum under consideration
behaves as a diamagnetic medium, a result that should be compared with a 
previous one \cite{C-PFRT98} which exhibits the paramagnetic properties
of a fermionic vacuum. The obtained results provide in the simple 
context of a quantum charged scalar field some physical insights 
that can be useful in the treatment of more complicated
gauge groups and boundary conditions. 

\noindent
{\bf Acknowledgements}\hfill\break
We thank J. Rafelski for several insightful discussions on this subject.
We also acknowledge  A. A. Actor for useful conversations on the subject
with one of us (A. C. T.).
M. V. C.-P. and C. F. thanks CNPq (The National Research
Council of Brazil) and M. R. Negr\~ao thanks CAPES (Brazilian Council for
Graduate Training) for partial financial support.

\clearpage
\centerline{FIGURE CAPTION}

\noindent
{\bf Figure 1.} The permeability constant $\mu$ as a function of the
confining distance $a$ in units of Compton wavelength $1/m$.
\end{document}